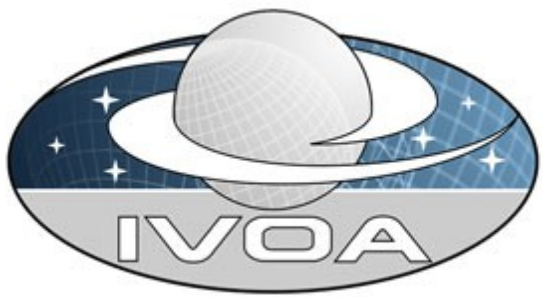

*International*

*Virtual*

*Observatory*

*Alliance*

# IVOA Simple Image Access
# Version 2.0
## IVOA Recommendation 2015-12-23

**Interest/Working Group:**

> http://www.ivoa.net/cgi-bin/twiki/bin/view/IVOA/IvoaDAL

**This version:**

> **http://www.ivoa.net/Documents/SIA/20151223/**

**Latest version:**

> http://www.ivoa.net/Documents/SIA/

**Previous version(s):**

> http://www.ivoa.net/Documents/SIA/20150730/
>
> http://www.ivoa.net/Documents/SIA/20150610/
>
> http://www.ivoa.net/Documents/SIA/20141024/
>
> http://www.ivoa.net/Documents/SIA/20140707/
>
> http://www.ivoa.net/Documents/SIA/20140512/
>
> http://www.ivoa.net/Documents/SIA/20131115/

**Editors:**

> **Patrick Dowler, François Bonnarel**

**Authors:**

> **Patrick Dowler, Doug Tody, François Bonnarel**

# Abstract

The Simple Image Access protocol (SIA) provides capabilities for the discovery, description, access, and retrieval of multi-dimensional *image* datasets, including 2-D images as well as datacubes of three or more dimensions.  SIA data discovery is based on the ObsCore Data Model (ObsCoreDM), which primarily describes data products by the physical axes (spatial, spectral, time, and polarization). Image datasets with dimension greater than 2 are often referred to as datacubes, *cube or image cube* datasets and may be considered examples of *hypercube* or *n-cube* data.  In this document the term "image" refers to general multi-dimensional datasets and is synonymous with these other terms unless the image dimensionality is otherwise specified.

SIA provides capabilities for image discovery and access. Data discovery and metadata access (using ObsCoreDM) are defined here. The capabilities for drilling down  to data files (and related resources) and services for remote access are defined elsewhere, but SIA also allows for direct access to retrieval.

# Status of This Document

This document has been produced by the Data Access Layer Working Group.

*It has been reviewed by IVOA Members and other interested parties, and has been endorsed by the IVOA Executive Committee as an IVOA Recommendation. It is a stable document and may be used as reference material or cited as a normative reference from another document. IVOA's role in making the Recommendation is to draw attention to the specification and to promote its widespread deployment. This enhances the functionality and interoperability inside the Astronomical Community.*

*A list of current IVOA Recommendations and other technical documents can be found at* [http://www.ivoa.net/Documents/](http://www.ivoa.net/Documents/).

# Acknowledgments

The authors would like to thank all the participants in DAL-WG discussions  for their ideas, critical reviews, and contributions to this document.

# Contents

# 1  Introduction

The Simple Image Access (SIA) protocol defines several capabilities to support discovery and access to astronomical image datasets of any dimension. Typical image datasets include 2-D spatial images, spectral data cubes, and cube and hypercube data of higher dimensions as well as derived image data products. The underlying ObsCore data model is a simplified view on the typical image datasets derived from observational data, which have some combination of spatial, spectral (including velocity and redshift), time, and polarization axes.

For complete access to datacubes, the SIA-2.0 specification  makes use of features defined in DataLink [8] . It also makes use of AccessData services, as well as custom data services.

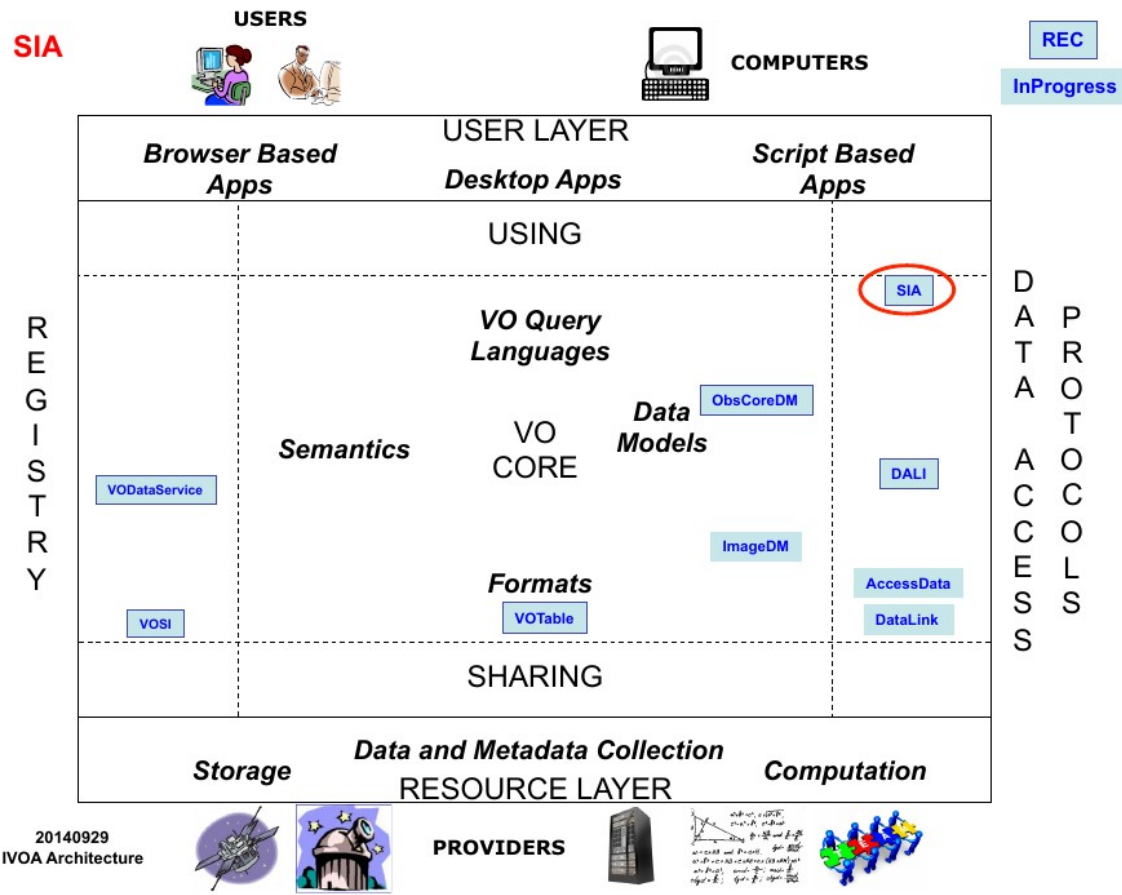

SIA defines data discovery and metadata capabilities that work with other DAL services to enable image and data cube access. The basic interface for the capabilities defined in this specification are described in DALI [1]. DataLink [2] can be used with SIA for finding access URL(s) for files, related resources, and data services such as AccessData (in development). SIA services also support VOSI-availability and VOSI-capabilities [2] resources.

The ObsCore data model has been defined in [7] it contains and organizes the minimal set of metadata necessary to discover datasets of interest for a specific purpose. The metadata returned from the SIA data discovery request is defined by the ObsCore [7] data model and serialized according to the ObsTAP specification [7]; this may be extended with additional metadata (columns)  in the future.  A future version of SIA may define a separate resource for accessing the complete  data model metadata for a single dataset once such a model is available. Data discovery responses are returned in VOTable [5] format unless an alternate format is requested.

## 1.1  The Role in the IVOA Architecture

SIA specifies standardID values for each capability, as defined by VODataService [10]. SIA services may be registered in an IVOA Registry using the SimpleDALRegExt [11] extension schema.

## 1.2  Changes from SIA-1.0 to SIA-2.0

Virtual Observatory access to astronomical images has been available via the SIA-1.0 protocol for over a decade.  Many such services have been implemented since 2002, and SIA-1.0 [3] was formally standardized as an IVOA Recommendation in 2009.  The legacy SIA standard however pre-dates much of the VO technology developed since 2002, and is limited to two-dimensional images.  SIA-2.0 is multi-dimensional and fully integrated with the modern VO architecture and related standards.

SIA-2.0 differs from legacy SIA-1.0 in the following aspects:

- The capabilities for dynamic access to image datasets are expanded in scope, but are separated from data discovery and download of whole image datasets.  A separate "AccessData" specification currently under development will define the more advanced dynamic data access functionality.  Automated virtual data generation and discovery (as in SIA-1.0) is not currently supported but is being considered for a future version of SIA.

- Description of Image datasets is now provided by the more abstract ObsCore data model [7], providing more comprehensive high level dataset metadata, most of which is common to other classes of astronomical data. Most of the attributes of the model can be queried using standard parameters, where SIA-1.0 only standardized positional queries.

- The version of VOTable [5]  used  in the protocol (2.1) has been updated and distinguishes between UCD and UType attributes. SIA-1.0 used a custom class of UCDs that pre-dated what are now UTypes, where ObsCore specifies standard UCD and UType values for use in the VOTable output.

- Most elements of the SIA-2.0 service interface are standardized across all DAL interfaces, as defined separately by the DALI [1] interface standard.

- SIAV2 supports the new DataLink [8] standard, providing additional capabilities for data access as well as the ability to link to related data products, as well as the VO Image data model standard currently under development, providing a full description of the structure of an image dataset.

While some of the more advanced capabilities for dynamic access to multi-dimensional images are still under development, the initial specification provides the capability to discover and download multi-dimensional image datasets via a service interface consistent with current VO standards

## 1.3  Current and Future support

These are the list of changes which  will be considered in a later version of this protocol  (SIA-2.1):

- The protocol should support  a detailed {metadata} capability. In the current design, this   capability will take a single parameter with an identifier (typically a publisher dataset identifier from a data discovery response) and return a document with detailed metadata about the dataset. The primary usage is to drill-down to the detailed  metadata after discovery

- The standardID for the {metadata} capability will  probably be : ivo://ivoa.net/std/SIA#metadata-2.1

- The parameter language should be extended to support case-insensitivity, wildcards, and pattern recognition.

- The DALI UPLOAD facility will be defined in order to allow queries based on values stored in a file.

- The protocol will also support an option for virtual data discovery. With this option, controlled by usage of an optional parameter, the  query can force the discovery of virtual datasets with best  matching to the input parameter constraints

- COLLECTION and FACILITY currently provide query parameters provide selection on service defined set of strings. It would be good to define standardized lists of those at IVOA level for next version.

- Visibility data and event lists are not considered as valid DPTYPES in this

version due to complexity of data access functionalities for those dataset types.

## 1.4 Motivating Use Cases

Below are some of the more common use cases that have motivated the development of the SIA specification. While this is not complete, it helps to understand the problem area covered by this specification.

### 1.4.1 Simple Data Discovery

Simple data discovery entails finding services that provide parameter based discovery of images and datacubes, querying the service(s) with a few well known kinds of queries that cover greater than 95% of use, and getting back easily parsed summary metadata about each available data product. The service discovery would be performed with an IVOA Registry search using a new service type defined for SIA-2.0.

The query for data would need to allow for querying in position, energy, time, and polarization:

- find data that includes specified coordinates (e.g. for some object)

- find data in the circle with coordinate centre and radius

- find data in a range of longitude and latitude

- find data within a specified simple polygon (one region, no holes, less than half the sphere)

- find data containing a specified energy (e.g. wavelength) or in a specified range of energy values

- find data obtained at a specified time (e.g. including a time instant) or during a specified range of times

- find data obtained with specified polarization (Stokes) states

- find data within a specified range of spatial resolution

- find data within a specified range of field-of-view

- find data within a range of exposure (integration) time

Queries can also combine any of these kinds of constraints (e.g. query using position *and* energy, position *and* time, etc.). Queries should be easily formulated with parameter-value pairs.

### 1.4.2  Get Detailed Metadata

The data discovery phase returns a subset of the available metadata.  Clients may need additional detailed metadata (as defined by the ImageDM specification) in order to make decisions or perform computations required to access the data (e.g. using a separate low-level data access service as described in the draft AccessData specification). The client must be able to easily figure out if detailed metadata is available and, using an identifier from the discovery response, make a call to a web service to retrieve the detailed metadata.

### 1.4.3  Download Complete Datasets

The client should be able to download complete datasets with information available in the discovery response. If the dataset is a single file, the service should provide an access URL to the file; if it is multiple files, then an access URL to a DataLink service [9] can be provided, but the client must be able to easily distinguish these two scenarios.

### 1.4.4  Access a Datacube with Operations: Too Big to Download

In many cases, datacubes are too large to download and process locally, so the client must be able to perform remote operations. Data discovery could be performed using any discovery protocol (SIA, TAP with ObsCore, etc.).  The client must be able to easily figure out if a low level access service is available for a discovered dataset. This could be using a URL provided in the response or by calling an associated DataLink service. Access operations include basic filtering (cut out a subsection of the data), transformations, or other pixel-level operations or even analysis. With current  version of the AccessData specification, we will only cover extracting a simple subset of an image or datacube.

## 1.5  Scope and Related Documents

This document can  support use cases 1.4.1, 1.4.3, and 1.4.4; support for 1.4.2 has been deferred to a future version. Some of the support for these use cases is provided by the separate capabilities defined in the DataLink [8] and AccessData specifications. Together, these three specifications, plus TAP [6], and  within the framework provided by ObsCore [7], and the future image and cube data model provide a set of capabilities required to support a broad range of use cases.

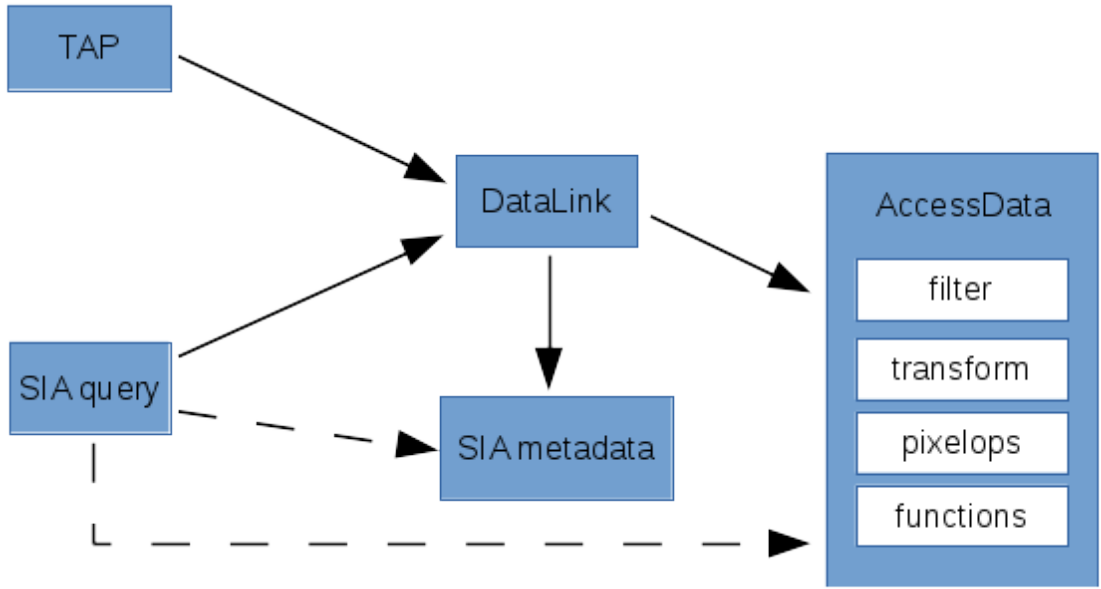

Each box in the above diagram shows a single capability. The SIA query capability is defined in this specification; the SIA metadata capability will be defined in a later version of this specification, once the ImageDM is completed. DataLink [8] and AccessData are separate specifications. The dashed lines represent optimisations that are mentioned in use cases above, where subsequent service usage should be easy to discover and invoke.

# 2 Resources

The SIA data discovery capability is implemented as a synchronous resource conforming to the DALI-sync [1] specification.

| resource type | resource name | required |
|---|---|---|
| {query} | service specific | yes |
| DALI-examples | /examples | no |
| VOSI-availability | /availability | yes |
| VOSI-capabilities | /capabilities | yes |

An SIA service must have at least one {query} resource; it could have multiple {query} resources (e.g. to support alternate authentication schemes where the path is different). All {query} resources must be siblings of the VOSI-capabilities resource; this limitation enables a client with just the URL for an SIA {query} resource (e.g. from a Datalink service descriptor) to find the VOSI-capabilities resource and discover all the capabilities provided.

## 2.1 {query} resource

The {query} resource is a synchronous web service resource that conforms to the DALI-sync description [1]. The implementer is free to name (set the path of) this resource however they like; the client will find the resource path using the VOSI-capabilities resource.

As a DALI-sync resource, the parameters for a request may be submitted using an HTTP GET (query string) or POST action.

All parameters for the {query} resource defined below **must** be supported by the service. Services must accept parameters and apply the constraints such that if a (ObsCore) record does not satisfy the constraints it is not included in the response. If the metadata for a field is not known (null), the constraint cannot be satisfied. The ObsCore data model [7] defines which fields may be null and which must have a value. For example, if dataset(s) have unknown time coverage ($t\_min$ and $t\_max$ in ObsCore), a query with the TIME parameter must not return the record(s); queries without the TIME constraint could still return such records, so the caller can discover such dataset(s).

Client requests may include zero or more of the query parameters.

All query parameters are **multi-valued which means multiple occurrences of the parameter=value pairs as specified in the DALI recommendation [1]** are permitted. The constraints from multiple occurrences of a parameter are combined with a logical OR operator. The constraints from different parameters are combined with a logical AND operator.

Query parameters for numeric fields accept a single floating point value or a range of values with optional lower and upper bounds. Such range values are encoded using the VOTable array serialisation (space separated). If the lower or upper bound is not specified, the range is open-ended. In VOTable arrays this uses the special values -Inf or +Inf. . For example, the interval [300,600] is:

```
300 600
```

The open-ended interval [300,infinity) (all values greater than or equal to 300) is:

```
300 +Inf
```

The open-ended interval (-infinity,600] (all values less than or equal to 600) is:

```
-Inf 600
```

The open-ended interval (-infinity,infinity) (all values) is:

```
-Inf +Inf
```

If specified, the boundary value is always included in the interval.

The units for numeric values are specified for each parameter and never included in the value.

Except where explicitly noted (see 2.1.10), query parameters for text or string fields are always case-sensitive and indicate an exact match.

The sections describing query parameters make use of fixed reference systems and units to simplify client and service implementation. These choices are not suitable for all domains; the values are chosen to enable the {query} resource to be used to search for most standard observational astronomy data. If they are not suitable for a specific domain of interest (e.g. planetary science) then it is feasible to write a very short standard that re-uses the SIA {query} capability but redefines the hard-coded systems and units. This new standard would have a new standardID to distinguish services that implement it from those that implement the capability defined here.

## 2.1.1  POS

The POS parameter defines the positional region(s) to be searched for data. The value is made up of a shape keyword followed by coordinate values. A POS constraint is satisfied if the specified shape intersects the bounds (s_region in the ObsCore [7] data model) of the observation.

The allowed shapes are:

| Shape | Coordinate values |
|---|---|
| CIRCLE | <longitude> <latitude> <radius> |
| RANGE | <longitude1> <longitude2>  <latitude1> <latitude2> |
| POLYGON | <longitude1> <latitude1> ... (at least 3 pairs) |

*Table 1: POS Values in Spherical Coordinates*

A circle at (12,34) with radius 0.5:

    POS=CIRCLE 12.0 34.0 0.5

A range of [12,14] in longitude and [34,36] in latitude:

    POS=RANGE 12.0 12.5 34.0 36.0

A polygon from (12,34) to (14,35) to (14,36) to (12,35) and (implicitly) back to (12,34):

    POS=POLYGON 12.0 34.0 14.0 35.0 14. 36.0 12.0 35.0

The inside is always assumed to be the smaller of the region to the left and the region to the right so only polygons smaller than half the sphere can be specified.

A band around the equator:

    POS=RANGE 0 360.0 -2.0 2.0

The north pole:

    POS=RANGE 0 360.0 89.0 +Inf

Although it is not really useful, the whole sky can be expressed:

    POS=RANGE -Inf +Inf -Inf +Inf

This syntax for circles and polygons is in the same style as STC-S, but with no reference positions, coordinate systems, units, or geometric operators (union, intersection, not). Coordinate values are floating point right ascension (RA) and declination (DEC) in ICRS and the units are always degrees. Valid coordinate values are in [0,360] for longitude and [-90,90] for latitude (or NaN).

## 2.1.2  BAND

The BAND parameter defines the energy interval(s) to be searched for data. The value is an open or closed energy interval. The interval always includes the bounding values. A BAND constraint is satisfied if the interval intersects the energy coverage of the observation ([em_min,em_max] in the ObsCore [7] data model).

Find data with spectral coverage that overlaps 500 to 550nm:

```
BAND=500e-9 550e-9
```

Find data with wavelength longer than 300m:

```
BAND=300 +Inf
```

Find data with wavelength shorter than 21cm:

```
BAND=-Inf 0.21
```

Find data that includes 21cm:

```
BAND=0.21
```

The scalar value 550, equivalent to [550,550]:

```
BAND=550
```

Searching using a scalar value should match any data that includes the specified value; searching using an interval will find any data with energy coverage that intersects the specified interval.

Energy values used in the BAND parameter are always assumed to be observed wavelength in meters. The ObsCore data model does not define a specific reference frame for values of em_min and em_max; values in the BAND parameter are assumed to be in the same (unspecified) frame as the content (e.g. no specific frame or transformation should be assumed).

### 2.1.3  TIME

The TIME parameter defines the time interval(s) to be searched for data. The value is an open or closed interval with numeric values (interpreted as Modified Julian Dates). A TIME constraint is satisfied if the interval intersects the time coverage of the observation ([t_min,t_max] in the ObsCore [7] data model).

A range of MJD values:

```
TIME=55123.456 55123.466
```

An instant in time:

```
TIME=55678.123456
```

Values used in the TIME parameter are always interpreted as specified in DALI [1] in the section on literal values: UTC time scale and UNKNOWN reference position [9]. Modified Julian Date values are always in days.

### 2.1.4 POL

The POL parameter defines the polarization state(s) to be searched for matching data.

Find data with unpolarized intensity:

```
POL=I
```

Find data with standard circular polarization:

```
POL=V
```

Find right or left circular polarized data:

```
POL=RR
POL=LL
```

Find data with any of IQU components:

```
POL=I
POL=Q
POL=U
```

The POL parameter constrains values of the pol_states column of the ObsCore data model [7]; possible values for the POL parameter are also defined by ObsCore [7].

### 2.1.5 FOV

The FOV parameter defines the range(s) of field of view (size) to be searched for data. This constraint is satisfied if the specified range includes the size of the field of view (s_fov column of the ObsCore [7] data model).

Find data with field of view between 1 and 2 degrees:

```
FOV=1.0 2.0
```

Find data with field of view larger than 1.0 degrees:

```
FOV=1.0 +Inf
```

Find data with field of view smaller than ~1 arcmin:

```
FOV=-Inf 0.017
```

Find data with very small (< 0.01 deg) or very large (> 2 deg) field of view:

```
FOV=-Inf 0.01
FOV=2.0 +Inf
```

Values used in the FOV parameter are always interpreted as angles expressed in degrees (same units as the s_fov column in the ObsCore [7] data model).

### 2.1.6 SPATRES

The SPATRES parameter define the range(s) of spatial resolution to be searched for data. This constraint is satisfied if the specified range includes the spatial resolution of the data (s_resolution column of the ObsCore [7] data model).

Find data with resolution better than 0.2 arcsec:

```
SPATRES=-Inf 0.2
```

Find data with resolution larger than 1.0 arcsec:

```
SPATRES=1.0 +Inf
```

Find data with resolution between 0.1 and 0.2 arcsec:

```
SPATRES=0.1 0.2
```

Values used in the SPATRES parameter are always interpreted as angles expressed in arcsec (same units as the s_resolution column in the ObsCore [7] data model).

### 2.1.7 SPECRP

The SPECRP parameter define the range(s) of spectral resolving power to be searched for data. This constraint is satisfied if the specified range includes the resolving power of the data (em_res_power column of the ObsCore [7] data model).

Find data with resolution better than 1000:

```
SPECRP=1000 +Inf
```

Find data with resolution less than 500:

```
SPECRP=-Inf 500
```

Find data with resolution between 10000 and 20000:

```
SPECRP=10000 20000
```

Values used in the SPECRP parameter are dimensionless ($\lambda/\Delta\lambda$, as for the em_res_power column in the ObsCore [7] data model).

### 2.1.8 EXPTIME

The EXPTIME parameter defines the range(s) of exposure times to be searched for data. This constraint is satisfied if the specified range includes the exposure time of the data (t_exptime column of the ObsCore [7] data model).

Find data with exposure time less than  60 seconds:

```
EXPTIME=-Inf 60
```

Find data with exposure time longer than 10 minutes:

```
EXPTIME=600 +Inf
```

Find data with exposure time between 10 and 30 minutes:

```
EXPTIME=600 1800
```

Find data with very short (< 2 sec) or very long (> 20 min) exposure time:

```
EXPTIME=-Inf 2
EXPTIME=1200 +Inf
```

Values used in the EXPTIME parameter are always expressed in seconds (same units as the t_exptime column in ObsCore).

### 2.1.9 TIMERES

The TIMERES parameter define the range(s) of temporal resolution to be searched for data. This constraint is satisfied if the specified range includes the temporal resolution of the data (t_resolution column of the ObsCore [7] data model).

Find data with resolution better than 1.0 sec:

```
TIMERES=-Inf 1.0
```

Find data with resolution larger than 1.0 sec:

```
TIMERES=1.0 +Inf
```

Find data with resolution between 1.0 and 2.0 sec:

```
TIMERES=1.0 2.0
```

Values used in the TIMERES parameter are always expressed in seconds (same units as the t_resolution column in the ObsCore [7] data model).

### 2.1.10 ID

The ID parameter is a string-valued parameter that specifies the identifier of dataset(s). Values of the ID parameter are compared to the obs_publisher_did column of the ObsCore [7] data model. Note that IVOIDs MUST be compared case-insensitively. As publisher dataset identifiers in the VO generally are IVOIDs, implementations will usually have to use case-insensitive comparisons here.

### 2.1.11 COLLECTION

The COLLECTION parameter is a string-valued parameter that specifies the name of the data collection. The value is compared with the obs_collection from the ObsCore [7] data model.

### 2.1.12 FACILITY

The FACILITY parameter is a string-valued parameter that specifies the name of the facility (usually telescope) where the data was acquired. The value is compared with the facility_name from the ObsCore [7] data model.

### 2.1.13 INSTRUMENT

The INSTRUMENT parameter is a string-valued parameter that specifies the name of the instrument with which the data was acquired. The value is compared with the instrument_name from the ObsCore [7] data model.

### 2.1.14 DPTYPE

The DPTYPE parameter is a string-valued parameter that specifies the type of data. The value is compared with the dataproduct_type from the ObsCore [7] data model. For the SIA {query} resource, the only values that should be returned for dataproduct_type are *image* and *cube*, so this parameter can be only really be used to select one of these.

### 2.1.15 CALIB

The CALIB parameter is a integer-valued parameter that specifies the calibration level of the data. The value is compared with the calib_level from the ObsCore [7] data model. To find raw data:

```
CALIB=0
CALIB=1
```

To find calibrated data:

```
CALIB=2
```

To find calibrated data and more highly processed data products:

```
CALIB=2
CALIB=3
```

### 2.1.16 TARGET

The TARGET parameter is a string-valued parameter that specifies the name of the target (e.g. the intention of the original science program or observation). The value is compared with the target_name from the ObsCore [7] data model.

### 2.1.17 FORMAT

The FORMAT parameter specifies response format(s) (the access_format column from the ObsCore [7] data model). This column describes the format of the response from the access_url (see 3.1.3) so the values could be data file types (e.g. application/fits) or they could be the DataLink [8] MIME type.

### 2.1.18 MAXREC

The MAXREC parameter is defined in DALI [1] and allows the client to limit the number or records in the response. A service implementation may also impose default and maximum values for this limit. However the limit is determined, if the output is truncated due to the limit the server must indicate this using an overflow (section 3.1) indicator except in the the special case of MAXREC=0, where the service respond with metadata-only (normal output document with no records).

### 2.1.19 UPLOAD

The DALI UPLOAD parameter is not used by this version of SIA. The use case of uploading lists of coordinates is covered by the multiple-valued parameters values.

### 2.1.20 Service PARAMETER self description

Any service may include a DataLink [8] service descriptor in the VOTable output to describe itself. This descriptor would describe the supported query parameters (standard and custom), including list of values for those with a fixed list (e.g. COLLECTION, INSTRUMENT, FACILITY, DPTYPE, CALIB, and FORMAT).

## 2.2 Availability: VOSI-availability

A web service with SIA capabilities must have a VOSI-availability resource [2] as described in DALI [1].

## 2.3 Capabilities: VOSI-capabilities

A web service with SIA capabilities must have a VOSI-capabilities resource [2] as described in DALI [1]. The standardID for the {query} capability is

```
ivo://ivoa.net/std/SIA#query-2.0
```

All DAL services must implement the *capabilities* resource. The following capabilities document shows the minimal metadata and does not require a registry extension schema:

```
<?xml version="1.0" encoding="UTF-8"?>
<vosi:capabilities
    xmlns:vosi="http://www.ivoa.net/xml/VOSICapabilities/v1.0"
    xmlns:xsi="http://www.w3.org/2001/XMLSchema-instance"
    xmlns:vs="http://www.ivoa.net/xml/VODataService/v1.1">
  <capability standardID="ivo://ivoa.net/std/VOSI#capabilities">
    <interface xsi:type="vs:ParamHTTP" version="1.0">
      <accessURL use="base">
        http://example.com/sia2/capabilities
      </accessURL>
    </interface>
  </capability>
  <capability standardID="ivo://ivoa.net/std/VOSI#availability">
    <interface xsi:type="vs:ParamHTTP" version="1.0">
      <accessURL use="full">
        http://example.com/sia2/availability
      </accessURL>
    </interface>
  </capability>
  <capability standardID="ivo://ivoa.net/std/SIA#query-2.0">
    <interface xsi:type="vs:ParamHTTP" role="std" version="2.0">
      <accessURL>
          http://example.com/sia2/query
      </accessURL>
    </interface>
    <!-- service details from extension schema could go here -->
  </capability>
</vosi:capabilities>
```

Note that the {query} resource does not have to be named as shown in the
accessURL(s) above. Multiple capability elements for the {query} and the
{metadata} resources may be included; this is typically used if they differ in
protocol (http vs. https) and/or authentication requirements.

# 3 {query} response

## 3.1 Successful Query

The response from a successful call to the {query} resource is a table consistent with ObsTAP responses as described in [7]. The ObsCore data model specifies all the (VOTable) field names, utypes, UCDs, and units to use in the response, as well as which fields must have values and which are allowed to be empty (null). The {query} response must contain the required ObsCore fields and may contain additional fields (from [7] or custom fields from the service provider). Examples are provided below (section 4).

Successfully executed requests should result in a response with HTTP status code 200 (OK) and a response in the format requested by the client or in the default format for the service. The default output format is VOTable. Other output formats can be specified by the RESPONSEFORMAT parameter (see [1]).

The service should set the following HTTP headers to the correct values where possible.

| Content-Type | mime-type of the response |
|---|---|
| Content-Encoding | encoding/compression of the response (if applicable) |

Table 2: Recommended HTTP Response Headers

Since the {query} response is usually dynamically generated, the Content-Length and Last-Modified headers cannot usually be set.

### 3.1.1 Related Service Metadata

The DataLink specification [9] gives a recipe for including additional resources in VOTable that enable the client to invoke services using values from the table as parameter values.

If the provider implements a DataLink service for the data being found via SIA, the {query} response **should** include a description for invoking the DataLink service, usually using values from the obs_publisher_did column.

If the provider implements an AccessData capability for the data being found via SIA and this capability can be invoked directly using an identifier in the {query} response, the {query} response **should** include a description for invoking this capability, usually using values from the obs_publisher_did column.

### 3.1.2 SIA {query} Service Descriptor

The DataLink [8] specification describes a mechanism for describing a service within a VOTable resource and recommends that services can describe themselves with a special resource with name="this". SIA {query} responses should include a descriptor describing both standard and custom query parameters (if applicable). The descriptor for a service with standard parameters (see 2.1) would be:

```
<RESOURCE type="meta" utype="adhoc:service" name="this">

 <PARAM name="standardID" datatype="char" arraysize="*"
        value="ivo://ivoa.net/std/SIA#query-2.0" />

 <PARAM name="accessURL" datatype="char" arraysize="*"
        value="http://example.com/sia2/query" />

 <GROUP name="inputParams">
  <PARAM name="POS" datatype="char" arraysize="*" xtype="circle" />
  <PARAM name="POS" datatype="char" arraysize="*" xtype="range" />
  <PARAM name="POS" datatype="char" arraysize="*" xtype="polygon" />
  <PARAM name="BAND" datatype="double" arraysize="*"
         xtype="interval" unit="m" />
  <PARAM name="TIME" datatype="double" arraysize="*"
         xtype="interval" unit="d" />
  <PARAM name="POL" datatype="char" arraysize="*" />
  <PARAM name="FOV" datatype="double" arraysize="*"
         xtype="interval" unit="deg" />
  <PARAM name="SPATRES" datatype="double" arraysize="*"
         xtype="interval" unit="arcsec" />
  <PARAM name="EXPTIME" datatype="double" arraysize="*"
         xtype="interval" unit="sec" />
  <PARAM name="ID" datatype="char" arraysize="*" />
  <PARAM name="COLLECTION" datatype="char" arraysize="*" />
  <PARAM name="FACILITY" datatype="char" arraysize="*" />
  <PARAM name="INSTRUMENT" datatype="char" arraysize="*" />
  <PARAM name="DPTYPE" datatype="char" arraysize="*" />
  <PARAM name="CALIB" datatype="int" />
  <PARAM name="TARGET" datatype="char" arraysize="*" />
  <PARAM name="TIMERES" datatype="double" arraysize="*"
         xtype="interval" unit="sec" />
  <PARAM name="SPECRP" datatype="double" arraysize="*"
         xtype="interval" />
  <PARAM name="FORMAT" datatype="char" arraysize="*" />
 </GROUP>
</RESOURCE>
```

This VOTable resource should be included in the output from all queries; it is especially useful for MAXREC=0 queries since inclusion of the self descriptor would mean that all inputs and outputs would be fully described.

### 3.1.3  Use of access_url and access_format

If the SIA service is only dealing with simple data (one file per result), the access_url column **may** be a link directly to that file, in which case the access_format column **should** specify the file format (e.g. application/fits).

If the data provider implements a DataLink service for the data being found via the SIA {query} capability, they **may** put a URL to invoke the DataLink {links} capability (with ID parameter and value) in the access_url column; if they do this, they **must** also put the standard DataLink MIME type [9] in the access_format column.

## 3.2  Errors

The error handling specified for DALI-sync resources applies to service failure. If the requested format is VOTable, the error document must be VOTable as described by DALI [1], except when the service is broken.  If the requested format is one of the plain text (csv or tsv) formats, the error document should also be plain text using the text/plain content-type.

The error message must start with one of the strings in the following table, in order of specificity:

| Error | Meaning |
|---|---|
| UsageFault | Invalid input (e.g. invalid input parameter value) |
| TransientFault | Service is not currently able to function |
| FatalFault | Service cannot perform requested action |
| DefaultFault | General error (not covered above) |

*Table 3: Error Messages*

In all cases, the service may append additional useful information to the error strings above. If there is additional text, it must be separated from the error string with a colon (:) character, for example:

```
UsageFault: invalid BAND value -2
```

# 4  Examples

This section presents two examples of queries and responses corresponding to the following scenarios :

How do I query a SIAV2 service containing  IRAS-IRIS images in a circle of 0.1 deg around position 2.8425 +74.4846 selecting 200 and 60 micron bands ?

Note: Spaces in parameter values must be URL-encoded as %2B or +; we leave this out of the example to make it easier to read.

http://dalservices.ivoa.net/sia2/query?POS=CIRCLE 2.8425 74.4846 0.1
&BAND=0.0002&BAND=0.00006&COLLECTION=IRAS-IRIS

```
<?xml version="1.0" encoding="UTF-8" ?>
<VOTABLE ersion="1.2" xmlns:xsi = "http://www.w3.org/2001/XMLSchema-instance" xsi:noNamespaceSchemaLocation =
"xmlns:http://www.ivoa.net/xml/VOTable-1.2.xsd" >
   <RESOURCE type="results">
   <INFO name="QUERY_STATUS" value="OK"/>
   <TABLE>
    <FIELD name="dataproduct_type" ucd="meta.id" datatype="char"
          utype="obscore:ObsDataSet.dataProductType" arraysize="*">
       <DESCRIPTION>Data product type</DESCRIPTION>
    </FIELD>
    <FIELD name="calib_level" ucd="meta.code;obs.calib" datatype="int" utype="obscore:ObsDataSet.calibLevel">
       <DESCRIPTION>Calibration level</DESCRIPTION>
    </FIELD>
    <FIELD name="obs_collection" datatype="char" ucd="meta.id" utype="obscore:DataID.Collection" arraysize="*">
       <DESCRIPTION>Data collection to which dataset belongs</DESCRIPTION>
    </FIELD>
    <FIELD name="obs_id" ucd="meta.id" datatype="char" utype="obscore:DataID.observationID" arraysize="*">
         <DESCRIPTION>Free syntax Observation Identifier</DESCRIPTION>
    </FIELD>
    <FIELD name="obs_publisher_did" ucd="meta.ref.url;meta.curation" datatype="char"
          utype="obscore:Curation.PublisherDID" arraysize="*">
      <DESCRIPTION>Publisher's ID for the dataset ID</DESCRIPTION>
    </FIELD>
    <FIELD name="access_url" ucd="meta.ref.url"  datatype="char" utype="obscore:Access.Reference" arraysize="*">
       <DESCRIPTION>URL used to access dataset</DESCRIPTION>
    </FIELD>
    <FIELD name="access_format" datatype="char" ucd="meta.code.mime" utype="obscore:Access.Format" arraysize="*">
       <DESCRIPTION>Content or MIME type of dataset</DESCRIPTION>
    </FIELD>
    <FIELD name="access_estsize" datatype="int" ucd="phys.size;meta.file" utype="obscore:Access.Size">
            <DESCRIPTION>Dataset estimated size</DESCRIPTION>
    </FIELD>
    <FIELD name="target_name" datatype="char" ucd="meta.id;src" utype="obscore:Target.Name" arraysize="*">
          <DESCRIPTION>Target name</DESCRIPTION>
    </FIELD>
    <FIELD name="s_ra" datatype="double" ucd="pos.eq.ra"
utype="obscore:Char.SpatialAxis.Coverage.Location.Coord.Position2D.Value2.C1" unit="deg" >
          <DESCRIPTION>Spatial Position RA</DESCRIPTION>
    </FIELD>
    <FIELD name="s_dec" datatype="double" ucd="pos.eq.dec"
          utype="obscore:Char.SpatialAxis.Coverage.Location.Coord.Position2D.Value2.C2" unit="deg" >
       <DESCRIPTION>Spatial Position Dec</DESCRIPTION>
    </FIELD>
    <FIELD name="s_fov" datatype="char" ucd="phys.angSize;instr.fov"
          utype="obscore:SpatialAxis.Coverage.Bounds.Extent.diameter" unit="deg" >
       <DESCRIPTION>Spatial Field of view "diameter"</DESCRIPTION>
    </FIELD>
    <FIELD name="s_region" datatype="char" ucd="phys.angArea;obs"
          utype="obscore:Char.SpatialAxis.Coverage.Support.Area" arraysize="*" unit="deg" >
          <DESCRIPTION>Spatial support</DESCRIPTION>
```

```
        </FIELD>
        <FIELD name="s_resolution" datatype="double" ucd="pos.angResolution"
              utype="obscore:Char.SpatialAxis.Resolution.refval.value" >
            <DESCRIPTION>Spatial resolution FWHM</DESCRIPTION>
        </FIELD>
        <FIELD name="t_min" datatype="double" ucd="time.start;obs.exposure"
              utype="obscore:Char.TimeAxis.Coverage.Bounds.Limits.StartTime" unit="s" >
             <DESCRIPTION>Time coordinate Lower limit</DESCRIPTION>
        </FIELD>
        <FIELD name="t_max" datatype="double" ucd="time.end;obs.exposure"
              utype="obscore:Char.TimeAxis.Coverage.Bounds.Limits.StopTime" unit="s">
             <DESCRIPTION>Time coordinate Higher limit</DESCRIPTION>
        </FIELD>
        <FIELD name="t_exptime" ucd="time.duration;obs.exposure" datatype="double"
              utype="obscore:Char.TimeAxis.Coverage.Support.Extent" unit="s" >
               <DESCRIPTION>Exposure time</DESCRIPTION>
        </FIELD>
        <FIELD name="t_resolution" datatype="double" ucd="time.resolution"
              utype="obscore:Char.TimeAxis.Resolution.refval.value" unit="s" >
               <DESCRIPTION>Time resolution</DESCRIPTION>
        </FIELD>
        <FIELD name="em_min" datatype="double" ucd="em.wl;stat.min"
                utype="obscore:Char.SpectralAxis.Coverage.Bounds.Limits.LoLimit" unit="m" >
               <DESCRIPTION>Spectral coordinate Lower limit</DESCRIPTION>
        </FIELD>
        <FIELD name="em_max" datatype="double" ucd="em.wl;stat.max"
              utype="obscore:Char.SpectralAxis.Coverage.Bounds.Limits.HiLimit"  unit="m">
               <DESCRIPTION>Spectral coordinate Higher limit</DESCRIPTION>
        </FIELD>
        <FIELD name="em_res_power" datatype="double" ucd="spect.resolution"
              utype="obscore:Char.SpectralAxis.Coverage.Resolution.ResolPower.refval" >
            <DESCRIPTION>SPECTRAL Resolving power</DESCRIPTION>
        </FIELD>
        <FIELD name="o_ucd" datatype="char" ucd="meta.ucd" utype="obscore:Char.ObservableAxis.ucd" arraysize="*"  >
               <DESCRIPTION>UCD specifying the quantity on Observable axis</DESCRIPTION>
        </FIELD>
        <FIELD name="pol_states" datatype="char" ucd="meta.code;phys.polarization"
              utype="obscore:Char.PolarizationAxis.stateList" arraysize="*" >
             <DESCRIPTION>Enumeration of Polarization sates</DESCRIPTION>
        </FIELD>
        <FIELD name="facility_name" datatype="char" ucd="meta.id;instr.tel"
utype="obscore:Provenance.ObsConfig.facility.name" arraysize="*">
               <DESCRIPTION>Facility name</DESCRIPTION>
        </FIELD>
        <FIELD name="instrument_name" ucd="meta.id;instr" datatype="char" arraysize="*"
              utype="obscore:Provenance.ObsConfig.instrument.name">
            <DESCRIPTION>Instrument name</DESCRIPTION>
        </FIELD>
        <DATA>
            <TABLEDATA>
                <TR>
                    <TD>cube</TD>
                    <TD>1</TD>
                    <TD>IRAS-IRIS</TD>
                    <TD>I422B2H0</TD>
                    <TD>ivo://cds.u-strasbg.fr/IRAS-IRIS/25MJ/I422B2H0</TD>
                    <TD><![CDATA[http://aladix.u-strasbg.fr/cgi-bin/nph-Aladin++dev.cgi?
out=image&position=0.000000+80.000000&field=I422B2H0&survey=IRAS-IRIS&color=25MU&mode=view]]></TD>
                    <TD>image/fits</TD>
                    <TD>1600</TD>
                    <TD>I422B2H0</TD>
                    <TD>0.000000 </TD>
                    <TD>80.000000 </TD>
                    <TD>0.5</TD>
                    <TD>POLYGON 30.0 200.0 32.0 200.0 32.0 198.0 30.0 198.0</TD>
                    <TD></TD>
                    <TD></TD>
                    <TD></TD>
```

```
                    <TD>1000</TD>
                    <TD>1.0</TD>
                    <TD>0.21</TD>
                    <TD>0.21</TD>
                    <TD>5.0</TD>
                    <TD></TD>
                    <TD>Stokes</TD>
                    <TD>IRAS-IRIS</TD>
                    <TD></TD>
                </TR>
                <TR>
                    <TD>cube</TD>
                    <TD>1</TD>
                    <TD>IRAS-IRIS</TD>
                    <TD>I408B1H0</TD>
                    <TD>ivo://cds.u-strasbg.fr/IRAS-IRIS/12MU/I408B1H0</TD>
                    <TD><![CDATA[http://aladix.u-strasbg.fr/cgi-bin/nph-Aladin++dev.cgi?
out=image&position=0.000000+70.000000&field=I408B1H0&survey=IRAS-IRIS&color=12MU&mode=view]]></TD>
                    <TD>image/fits</TD>
                    <TD>1600</TD>
                    <TD>I408B1H0</TD>
                    <TD>0.000000 </TD>
                    <TD>70.000000 </TD>
                    <TD>0.5</TD>
                    <TD>POLYGON ICRS 30.0 200.0 32.0 200.0 32.0 198.0 30.0 198.0</TD>
                    <TD></TD>
                    <TD></TD>
                    <TD></TD>
                    <TD>1000</TD>
                    <TD>1.0</TD>
                    <TD>0.21</TD>
                    <TD>0.21</TD>
                    <TD>5.0</TD>
                    <TD></TD>
                    <TD>Stokes</TD>
                    <TD>IRAS-IRIS</TD>
                    <TD></TD>
                </TR>
                <TR>
                    <TD>cube</TD>
                    <TD>1</TD>
                    <TD>IRAS-IRIS</TD>
                    <TD>I422B1H0</TD>
                    <TD>ivo://cds.u-strasbg.fr/IRAS-IRIS/12MU/I422B1H0</TD>
                    <TD><![CDATA[http://aladix.u-strasbg.fr/cgi-bin/nph-Aladin++dev.cgi?
out=image&position=0.000000+80.000000&field=I422B1H0&survey=IRAS-IRIS&color=12MU&mode=view]]></TD>
                    <TD>image/fits</TD>
                    <TD>1600</TD>
                    <TD>I422B1H0</TD>
                    <TD>0.000000 </TD>
                    <TD>80.000000 </TD>
                    <TD>0.5</TD>
                    <TD>POLYGON ICRS 30.0 200.0 32.0 200.0 32.0 198.0 30.0 198.0</TD>
                    <TD></TD>
                    <TD></TD>
                    <TD></TD>
                    <TD>1000</TD>
                    <TD>1.0</TD>
                    <TD>0.21</TD>
                    <TD>0.21</TD>
                    <TD>5.0</TD>
                    <TD></TD>
                    <TD>Stokes</TD>
                    <TD>IRAS-IRIS</TD>
                    <TD></TD>
                </TR>
            </TABLEDATA>
        </DATA>
```

```
        </TABLE>
      </RESOURCE>
</VOTABLE>
```

How do I query a service containing radio cubes from Alma operated by NRAO in a circle of 0.1 deg around position 180.47 -18.70 in the CO band in the range 800 microns to 900 microns and made in the time range between Mjd=55708 and 55710 ?

Note: Spaces in parameter values must be URL-encoded as %2B or +; we leave this out of the example to make it easier to read.

http://dalservices.ivoa.net/sia_b?REQUEST=query&POS=CIRCLE 180.475 -18.70 0.01&BAND= 0.0008 0.0009&TIME= 55708 55710&COLLECTION=ALMA

```
<?xml version="1.0" encoding="UTF-8" ?>

    <VOTABLE version="1.2" xmlns:xsi = "http://www.w3.org/2001/XMLSchema-instance" xsi:noNamespaceSchemaLocation =
"xmlns:http://www.ivoa.net/xml/VOTable-1.2.xsd" >
      <DESCRIPTION>DALServer SIAP Version 2.0 (vao-fall2013)</DESCRIPTION>
      <PARAM name="SORTORDER" datatype="char" value="Creator " />
      <RESOURCE type="results">
      <INFO name="QUERY_STATUS" value="OK"/>
      <TABLE>
      <FIELD name="dataproduct_type" ucd="meta.id" datatype="char"
utype="obscore:ObsDataSet.dataProductType" arraysize="*">
      <DESCRIPTION>Data product type</DESCRIPTION>
      </FIELD>
      <FIELD name="calib_level" ucd="meta.code;obs.calib" datatype="int" utype="obscore:ObsDataSet.calibLevel">
       <DESCRIPTION>Calibration level</DESCRIPTION>
      </FIELD>
      <FIELD name="obs_collection" datatype="char" ucd="meta.id" utype="obscore:DataID.Collection" arraysize="*">
        <DESCRIPTION>Data collection to which dataset belongs</DESCRIPTION>
      </FIELD>
      <FIELD name="obs_id" ucd="meta.id" datatype="char" utype="obscore:DataID.observationID" arraysize="*">
         <DESCRIPTION>Free syntax Observation Identifier</DESCRIPTION>
      </FIELD>
      <FIELD name="obs_publisher_did" ucd="meta.ref.url;meta.curation" datatype="char"
         utype="obscore:Curation.PublisherDID" arraysize="*">
       <DESCRIPTION>Publisher's ID for the dataset ID</DESCRIPTION>
      </FIELD>
      <FIELD name="access_url" ucd="meta.ref.url"  datatype="char" utype="obscore:Access.Reference" arraysize="*">
        <DESCRIPTION>URL used to access dataset</DESCRIPTION>
      </FIELD>
      <FIELD name="access_format" datatype="char" ucd="meta.code.mime" utype="obscore:Access.Format" arraysize="*">
        <DESCRIPTION>Content or MIME type of dataset</DESCRIPTION>
      </FIELD>
      <FIELD name="access_estsize" datatype="int" ucd="phys.size;meta.file" utype="obscore:Access.Size">
           <DESCRIPTION>Dataset estimated size</DESCRIPTION>
      </FIELD>
      <FIELD name="target_name" datatype="char" ucd="meta.id;src" utype="obscore:Target.Name" arraysize="*">
           <DESCRIPTION>Target name</DESCRIPTION>
      </FIELD>
      <FIELD name="s_ra" datatype="double" ucd="pos.eq.ra"
utype="obscore:Char.SpatialAxis.Coverage.Location.Coord.Position2D.Value2.C1" unit="deg" >
           <DESCRIPTION>Spatial Position RA</DESCRIPTION>
      </FIELD>
      <FIELD name="s_dec" datatype="double" ucd="pos.eq.dec"
           utype="obscore:Char.SpatialAxis.Coverage.Location.Coord.Position2D.Value2.C2" unit="deg" >
        <DESCRIPTION>Spatial Position Dec</DESCRIPTION>
      </FIELD>
      <FIELD name="s_fov" datatype="char" ucd="phys.angSize;instr.fov"
           utype="obscore:SpatialAxis.Coverage.Bounds.Extent.diameter" unit="deg" >
        <DESCRIPTION>Spatial Field of view "diameter"</DESCRIPTION>
      </FIELD>
      <FIELD name="s_region" datatype="char" ucd="phys.angArea;obs"
           utype="obscore:Char.SpatialAxis.Coverage.Support.Area" arraysize="*" unit="deg" >
```

```xml
<DESCRIPTION>Spatial support</DESCRIPTION>
        </FIELD>
        <FIELD name="s_resolution" datatype="double" ucd="pos.angResolution"
              utype="obscore:Char.SpatialAxis.Resolution.refval.value" >
            <DESCRIPTION>Spatial resolution FWHM</DESCRIPTION>
        </FIELD>
        <FIELD name="t_min" datatype="double" ucd="time.start;obs.exposure"
              utype="obscore:Char.TimeAxis.Coverage.Bounds.Limits.StartTime" unit="s" >
             <DESCRIPTION>Time coordinate Lower limit</DESCRIPTION>
        </FIELD>
        <FIELD name="t_max" datatype="double" ucd="time.end;obs.exposure"
              utype="obscore:Char.TimeAxis.Coverage.Bounds.Limits.StopTime" unit="s">
             <DESCRIPTION>Time coordinate Higher limit</DESCRIPTION>
        </FIELD>
        <FIELD name="t_exptime" ucd="time.duration;obs.exposure" datatype="double"
              utype="obscore:Char.TimeAxis.Coverage.Support.Extent" unit="s" >
             <DESCRIPTION>Exposure time</DESCRIPTION>
        </FIELD>
        <FIELD name="t_resolution" datatype="double" ucd="time.resolution"
              utype="obscore:Char.TimeAxis.Resolution.refval.value" unit="s" >
             <DESCRIPTION>Time resolution</DESCRIPTION>
        </FIELD>
         <FIELD name="em_min" datatype="double" ucd="em.wl;stat.min"
                utype="obscore:Char.SpectralAxis.Coverage.Bounds.Limits.LoLimit" unit="m" >
             <DESCRIPTION>Spectral coordinate Lower limit</DESCRIPTION>
        </FIELD>
        <FIELD name="em_max" datatype="double" ucd="em.wl;stat.max"
              utype="obscore:Char.SpectralAxis.Coverage.Bounds.Limits.HiLimit"  unit="m">
             <DESCRIPTION>Spectral coordinate Higher limit</DESCRIPTION>
        </FIELD>
        <FIELD name="em_res_power" datatype="double" ucd="spect.resolution"
              utype="obscore:Char.SpectralAxis.Coverage.Resolution.ResolPower.refval" >
            <DESCRIPTION>SPECTRAL Resolving power</DESCRIPTION>
        </FIELD>
        <FIELD name="o_ucd" datatype="char" ucd="meta.ucd" utype="obscore:Char.ObservableAxis.ucd" arraysize="*" >
             <DESCRIPTION>UCD specifying the quantity on Observable axis</DESCRIPTION>
        </FIELD>
        <FIELD name="pol_states" datatype="char" ucd="meta.code;phys.polarization"
              utype="obscore:Char.PolarizationAxis.stateList" arraysize="*" >
            <DESCRIPTION>Enumeration of Polarization sates</DESCRIPTION>
        </FIELD>
        <FIELD name="facilty_name" datatype="char" ucd="meta.id;instr.tel"
utype="obscore:Provenance.ObsConfig.facility.name" arraysize="*">
            <DESCRIPTION>Facility name</DESCRIPTION>
        </FIELD>
        <FIELD name="instrument_name" ucd="meta.id;instr" datatype="char" arraysize="*"
              utype="obscore:Provenance.ObsConfig.instrument.name">
            <DESCRIPTION>Instrument name</DESCRIPTION>
        </FIELD>

        <DATA>
          <TABLEDATA>
            <TR>
              <TD>image</TD>
              <TD>1</TD>
              <TD>ALMA_test</TD>
              <TD>ALMA test data: Antennae_South CO3_Line Full_velocity_map</TD>
              <TD>ivo://nrao/vo#siav2model:373</TD>
              <TD><![CDATA[http://vaosa-vm1.aoc.nrao.edu/ivoa-dal/siapv2-vao/sync?
REQUEST=accessData&FORMAT=image/fits&PubDID=ivo%3A%2F%2Fnrao%2Fvo%23siav2model%3A373]]></TD>
              <TD>image/fits</TD>
               <TD>2250000</TD>
              <TD></TD>
              <TD>180.475103</TD>
              <TD>-18.8855694</TD>
              <TD>0.027</TD>
```

```
            <TD>POLYGON 180.4607706  -18.8991286 180.4607706 -18.8720092  180.4894354 -18.8720092 180.4894354
-18.8991286</TD>
              <TD>3.6e-5</TD>
              <TD>55709.129226</TD>
            <TD>55709.129226</TD>
            <TD/>
            <TD/>
            <TD>0.00087103558671619</TD>
            <TD>0.000873082347755277</TD>
            <TD>29753.745800131965</TD>
            <TD>phot.flux</TD>
            <TD>I</TD>
            <TD>ALMA</TD>
            <TD/>
            </TR>
            <TR>
            <TD>image</TD>
            <TD>1</TD>
            <TD>ALMA_test</TD>
            <TD>ALMA test data: Antennae_South CO3_2Line  Velocity_map_cutout</TD>
            <TD>ivo://nrao/vo#siav2model:373</TD>
            <TD><![CDATA[http://vaosa-vm1.aoc.nrao.edu/ivoa-dal/siapv2-vao/sync?
REQUEST=accessData&FORMAT=image/fits&PubDID=ivo%3A%2F%2Fnrao%2Fvo%23image-ZSKuYR]]></TD>
            <TD>image/fits</TD>
            <TD>292456</TD>
            <TD></TD>
            <TD>180.474988373509</TD>
            <TD>-18.8799902066141</TD>
            <TD>0.0097</TD>
            <TD>POLYGON ICRS 180.4607706 -18.8991286 180.4607706 -18.8720092  180.4894354 -18.8720092 180.4894354
-18.8991286 </TD>
            <TD>3.6e-5</TD>
            <TD>55709.129226</TD>
            <TD>55709.129226</TD>
            <TD/>
            <TD/>
            <TD>0.00087410933963009</TD>
            <TD>0.000870797535056077</TD>
            <TD>29753.745800131965</TD>
            <TD>phot.flux</TD>
            <TD>I</TD>
            <TD>ALMA</TD>
            <TD/>
            </TR>
            <TR>
            <TD>image</TD>
            <TD>1</TD>
            <TD>ALMA_test</TD>
            <TD>ALMA test data: Antennae_South CO3_2Line Full_integrated_intensity</TD>
            <TD>ivo://nrao/vo#siav2model:374</TD>
            <TD><![CDATA[http://vaosa-vm1.aoc.nrao.edu/ivoa-dal/siapv2-vao/sync?
REQUEST=accessData&FORMAT=image/fits&PubDID=ivo%3A%2F%2Fnrao%2Fvo%23siav2model%3A374]]></TD>
            <TD>image/fits</TD>
            <TD>2250000</TD>
            <TD></TD>
            <TD>180.475103</TD>
            <TD>-18.8855694</TD>
            <TD>0.027</TD>
            <TD>POLYGON ICRS 180.4607706 -18.8991286 180.4607706 -18.8720092 180.4894354 -18.8720092 180.4894354
-18.8991286 </TD>
            <TD>3.6e-5</TD>
            <TD>55709.129226</TD>
            <TD>55709.129226</TD>
            <TD/>
            <TD/>
            <TD>0.00087410933963009</TD>
            <TD>0.00087103558671619</TD>
            <TD>29753.745800131965</TD>
```

```
                    <TD>phot.flux</TD>
                    <TD>I</TD>
                    <TD>ALMA</TD>
                    <TD/>
                    </TR>
                    <TR>
                    <TD>image</TD>
                    <TD>1</TD>
                    <TD>ALMA_test</TD>
                    <TD>ALMA test data: Antennae_South CO3_2Line  Full_data_cube</TD>
                    <TD>ivo://nrao/vo#siav2model:379</TD>
                    <TD><![CDATA[http://vaosa-vm1.aoc.nrao.edu/ivoa-dal/siapv2-vao/sync?
REQUEST=accessData&FORMAT=image/fits&PubDID=ivo%3A%2F%2Fnrao%2Fvo%23siav2model%3A379]]></TD>
                    <TD>image/fits</TD>
                    <TD>157500000</TD>
                    <TD></TD>
                    <TD>180.475103</TD>
                    <TD>-18.8855694</TD>
                    <TD>0.027</TD>
                    <TD>POLYGON ICRS 180.4607706 -18.8991286 180.4607706 -18.8720092 180.4894354 -18.8720092 180.4894354
-18.8991286 </TD>
                    <TD>3.6e-5</TD>
                    <TD>55709.129226</TD>
                    <TD>55709.129226</TD>
                    <TD></TD>
                    <TD></TD>
                    <TD>0.00087410933963009</TD>
                    <TD>0.000871035586716119</TD>
                    <TD>29753.745800131965</TD>
                    <TD>phot.flux</TD>
                    <TD>I</TD>
                    <TD>ALMA</TD>
                    <TD/>
                    </TR>
                    <TR>
                    <TD>image</TD>
                    <TD>1</TD>
                    <TD>ALMA_test</TD>
                    <TD>ALMA test data: Antennae_South CO3_2Line  Cube_cutout</TD>
                    <TD>ivo://nrao/vo#siav2model:379</TD>
                    <TD><![CDATA[http://vaosa-vm1.aoc.nrao.edu/ivoa-dal/siapv2-vao/sync?
REQUEST=accessData&FORMAT=image/fits&PubDID=ivo%3A%2F%2Fnrao%2Fvo%23image-m0RXVX]]></TD>
                    <TD>image/fits</TD>
                    <TD>20471920</TD>
                    <TD></TD>
                    <TD>180.474988373509</TD>
                    <TD>-18.8799902066141</TD>
                    <TD>0.0097</TD>
                    <TD>POLYGON ICRS 180.4607706 -18.8991286 180.4607706 -18.8720092 180.4894354 -18.8720092 180.4894354
-18.8991286</TD>
                    <TD>3.6e-5</TD>
                    <TD>55709.129226</TD>
                    <TD>55709.129226</TD>
                    <TD></TD>
                    <TD></TD>
                    <TD>0.00087410933963009</TD>
                    <TD>0.000872463699874965</TD>
                    <TD>29753.745800131965</TD>
                    <TD>phot.flux</TD>
                    <TD>I</TD>
                    <TD>ALMA</TD>
                    <TD/>
                    </TR>
                    <TR>
                    <TD>image</TD>
                    <TD>1</TD>
                    <TD>ALMA_test</TD>
                    <TD>ALMA test data: Antennae_South CO3_2Line  full_dispersion_map</TD>
```

```
            <TD>ivo://nrao/vo#siav2model:381</TD>
            <TD><![CDATA[http://vaosa-vm1.aoc.nrao.edu/ivoa-dal/siapv2-vao/sync?
REQUEST=accessData&FORMAT=image/fits&PubDID=ivo%3A%2F%2Fnrao%2Fvo%23siav2model%3A381]]></TD>
            <TD>image/fits</TD>
            <TD>2250000</TD>
            <TD></TD>
            <TD>180.475103</TD>
            <TD>-18.8855694</TD>
            <TD>0.027</TD>
            <TD>POLYGON ICRS 180.4607706 -18.8991286 180.4607706 -18.8720092 180.4894354 -18.8720092 180.4894354
-18.8991286</TD>
            <TD>3.6e-5</TD>
            <TD>55709.129226</TD>
            <TD>55709.129226</TD>
            <TD></TD>
            <TD/>
            <TD>0.000871035586716119</TD>
            <TD>0.000874109339630009</TD>
            <TD>29753.745800131965</TD>
            <TD>phot.flux</TD>
            <TD>I</TD>
            <TD>ALMA</TD>
            <TD/>
            </TR>
            <TR>
            <TD>image</TD>
            <TD>1</TD>
            <TD>ALMA_test</TD>
            <TD>ALMA test data:  Antennae_South.CO3_2Line.Clean.pcal1.image.mom.weighted_dispersion_coord.fits</TD>
            <TD>ivo://nrao/vo#siav2model:381</TD>
            <TD><![CDATA[http://vaosa-vm1.aoc.nrao.edu/ivoa-dal/siapv2-vao/sync?
REQUEST=accessData&FORMAT=image/fits&PubDID=ivo%3A%2F%2Fnrao%2Fvo%23image-ByCXy3]]></TD>
            <TD>image/fits</TD>
            <TD>292456</TD>
            <TD></TD>
            <TD>180.474988373509</TD>
            <TD>-18.8799902066141</TD>
            <TD>0.0097</TD>
            <TD>POLYGON ICRS 180.4607706 -18.8991286 180.4607706 -18.8720092 180.4894354 -18.8720092 180.4894354
-18.8991286</TD>
            <TD>3.6e-5</TD>
            <TD>55709.129226</TD>
            <TD>55709.129226</TD>
            <TD></TD>
            <TD></TD>
            <TD>0.000870797535056077</TD>
            <TD>0.000874109339630009</TD>
            <TD>29753.745800131965</TD>
            <TD>phot.flux</TD>
            <TD>I</TD>
            <TD>ALMA</TD>
            <TD/>
            </TR>
          </TABLEDATA>
        </DATA>
      </TABLE>
    </RESOURCE>
</VOTABLE>
```

# 5 Changes

## 5.1 REC-SIA-2.0-20151223

Fix typos.

## 5.2 PR-SIA-2.0-20151029

Changed open-ended interval bounds from NaN to -Inf or +Inf for consistency with VOTable double values.

Formatting fixes and regenerated table of contents.

## 5.3 PR-SIA-2.0-20150730

Editorial and organisational changes from TCG review.

## 5.4 PR-SIA-2.0-20150610

Changed the serialisation of numeric ranges in parameter values to use standard VOTable array format (space separated) so that parameters can be correctly described as accepting an array of numeric values rather than a string. Removed support for ISO8601 (timestamp) format from TIME parameter.

Added subsection for MAXREC parameter. Changed use of UNKNOWN (concept from STC) coordinate system to a plain text description. Clarified that numeric arguments to POS are floating point values (in examples and text).

Moved section comparing SIA-1.0 and SIA-2.0 earlier in the introduction.

Added example output for queries.

Add restriction that {query} resources must be a sibling of VOSI-capabilities.

Many minor clarifications and tweaks as recommended by TCG review.

## 5.5 PR-SIA-2.0-20141020

Added section to the introduction describing the conceptual difference between SIA1.0 and SIA-2.0 and related specifications.

Clarified that all text or string {query} parameters are case-sensitive in this initial version. Added valid ranges of coordinate values for POS parameter.

Added IVOA architecture document and replaced get-gory-details with "SIA metadata" in the DAL architecture diagram.

Fixed the status text to be the right text for PR.

Fixed numerous small typos and removed some cruft related to REQUEST.

Added some missing references.

### 5.6  PR-SIA-2.0-20140707

Promoted from WD to PR.

### 5.7  WD-SIA-2.0-20140707

Added query parameters for remaining useful ObsCore fields and clarified that all query parameters are required and how null values for ObsCore fields are treated in queries.

Removed REQUEST parameter, as per discussions at the interop.

Removed standard error message labels for authentication and authorization failures since these are difficult to implement consistently in different web service platforms. Changed the error message strings to use the word Fault (following DataLink and GWS-WG style) since Error has specific meaning in some platforms.

### 5.8  WD-SIA-2.0-20140505

FORMAT and UPLOAD parameters are still TBD.

Added FOV, SPATRES, and EXPTIME parameters to query ObsCore fields s_fov, s_resolution, and t_exptime respectively.

Removed the BOX from POS shape list.

Fixed BAND parameter to specify that values are vacuum wavelength in meters with UNKNOWN reference position.

Clarified TIME parameter to specify values are UTC time scale with UNKNOWN reference position (as in DALI).

Marked text related to the {metadata} capability as placeholder(s) that are only informational in the SIA-2.0 specification.

Removed the section on DALI-examples because it didn't add anything.

Added missing query constraints to the use cases in 1.2.1 (from CSP summary).

Clarified relationship between query parameters are fields/columns in the ObsCore data model.

Added standard error message strings.

### 5.9 WD-SIA-2.0

This is a major rewrite of previous drafts after several meetings, prototypes, and architectural changes. A real change log will begin with the next draft.